\documentclass[prb, aps, 10pt, superscriptaddress, twocolumn]{revtex4-1}
\usepackage[T1]{fontenc}
\usepackage[latin1]{inputenc}
\usepackage{latexsym}
\usepackage[pdftex,colorlinks=true, pdfstartview=FitH,linkcolor=blue, citecolor=red, urlcolor=green]{hyperref}
\usepackage{docs}
\usepackage{graphicx}	
\usepackage{amssymb}
\usepackage{amsmath}
\usepackage{bm}
\usepackage{subfigure}

\begin{document}

\title{Optoelectronic forces with quantum wells for cavity optomechanics in GaAs/AlAs semiconductor microcavities}

\author{V. Villafa\~ne}
\affiliation{Centro At\'omico Bariloche \& Instituto Balseiro (C.N.E.A.) and CONICET, 8400 S. C. de Bariloche, R. N., Argentina.}

\author{P. Sesin}
\affiliation{Centro At\'omico Bariloche \& Instituto Balseiro (C.N.E.A.) and CONICET, 8400 S. C. de Bariloche, R. N., Argentina.}

\author{P. Soubelet}
\affiliation{Centro At\'omico Bariloche \& Instituto Balseiro (C.N.E.A.) and CONICET, 8400 S. C. de Bariloche, R. N., Argentina.}

\author{S. Anguiano}
\affiliation{Centro At\'omico Bariloche \& Instituto Balseiro (C.N.E.A.) and CONICET, 8400 S. C. de Bariloche, R. N., Argentina.}

\author{A. E. Bruchhausen}
\affiliation{Centro At\'omico Bariloche \& Instituto Balseiro (C.N.E.A.) and CONICET, 8400 S. C. de Bariloche, R. N., Argentina.}

\author{G. Rozas}
\affiliation{Centro At\'omico Bariloche \& Instituto Balseiro (C.N.E.A.) and CONICET, 8400 S. C. de Bariloche, R. N., Argentina.}

\author{C. Gomez Carbonell}
\affiliation{Centre de Nanosciences et de Nanotechnologies, C.N.R.S., Univ. Paris-Sud, Universit\'e Paris-Saclay, C2N Marcoussis, 91460 Marcoussis, France.}

\author{A. Lema\^itre}
\affiliation{Centre de Nanosciences et de Nanotechnologies, C.N.R.S., Univ. Paris-Sud, Universit\'e Paris-Saclay, C2N Marcoussis, 91460 Marcoussis, France.}

\author{A. Fainstein}
\email[Corresponding author. E-mail: ]{afains@cab.cnea.gov.ar}
\affiliation{Centro At\'omico Bariloche \& Instituto Balseiro (C.N.E.A.) and CONICET, 8400 S. C. de Bariloche, R. N., Argentina.}

\date{\today}

%\ociscodes{(140.3490) Lasers, distributed feedback; (060.2420) Fibers, polarization-maintaining;(060.3735) Fiber Bragg gratings.}

%\doi{\url{http://dx.doi.org/10.1364/optica.XX.XXXXXX}}

\begin{abstract}
Radiation pressure, electrostriction, and photothermal forces have been investigated to evidence backaction, non-linearities and quantum phenomena in cavity optomechanics.  We show here through a detailed study of the relative intensity of the cavity mechanical modes observed when exciting with pulsed lasers close to the GaAs optical gap that optoelectronic forces involving real carrier excitation and deformation potential interaction are the strongest mechanism of light-to-sound transduction in semiconductor GaAs/AlAs distributed Bragg reflector optomechanical resonators. We demonstrate that the ultrafast spatial redistribution of the photoexcited carriers in microcavities with massive GaAs spacers leads to an enhanced coupling to the fundamental 20~GHz vertically polarized mechanical breathing mode. The carrier diffusion along the growth axis of the device can be enhanced by increasing the laser power, or limited by embedding GaAs quantum wells in the cavity spacer, a strategy used here to prove and engineer the optoelectronic forces in phonon generation with real carriers. The wavelength dependence of the observed phenomena provide further proof of the role of optoelectronic forces. The optical forces associated to the different intervening mechanisms and their relevance for dynamical backaction in optomechanics are evaluated using finite-element methods. The results presented open the path to the study of hitherto seldom investigated dynamical backaction in optomechanical solid-state resonators in the presence of optoelectronic forces.
\end{abstract}

\maketitle

\section{Motivation}

Backaction in cavity optomechanics has shown to lead to novel physical phenomena including laser cooling, self-oscillation, and non-linear dynamics in systems that go from kilometer size interferometers to single or few trapped ions.~\cite{ReviewCOM} Briefly, a resonant photon field in a cavity exerts a force and induces a mechanical motion on the mirrors, which in turn leads to a delayed modification of the resonant condition of the trapped field. Such coupled dynamics can be exploited for a large variety of applications that span for example from gravitational wave detection~\cite{Ligo} to the study of quantum motion states in mesoscopic mechanical systems.~\cite{O'Connell,Teufel,Chan,Verhagen} How light exerts force on matter is at the center of these investigations. Photons can apply stress through radiation pressure, transferring impulse when reflected from the mirrors.~\cite{Cohadon}  Related mechanisms also derived from the same fundamental interaction (Lorentz forces) are gradient forces (also exploited in optical tweezers)~\cite{Lin}, and electrostriction. The latter, linked to the material's photoelasticity, has been shown to play a role that can be of the same magnitude as radiation pressure,~\cite{Rakich1,Rakich2} or even larger if optical resonances are exploited in direct bandgap materials as for example GaAs.~\cite{FainsteinPRL2013,Rozas_Polariton,Baker}

In the presence of radiation pressure forces, the energy $E$ of the photon is shifted by the Doppler effect by an amount of the order $(v/c)E$, where $v$ and $c$ are the mirror and light velocities, respectively. The mechanical energy transferred from the photon to the mirror is thus very small. Electrostriction leads to Raman-like processes, for which the transferred energy $\Delta E_R$ corresponds to the involved vibrational frequency. This leads to inelastic scattering sidebands. Again, typically $\Delta E_R << E$. Contrastingly, if the photon is absorbed in the process of interaction, all its energy is transferred to the mirror. This fundamental difference has been used in cavity optomechanics to demonstrate strongly enhanced light-matter interactions based on photothermal forces.~\cite{MetzgerPRL2008,MetzgerPRB2008,Restrepo} In materials displaying optical resonances, the photons can be resonantly absorbed with the consequent transfer of electrons to excited states. Photoexcited carrier-mediated optomechanical interactions have been reported in semiconductor modulation-doped heterostructure-cantilever  hybrid systems. Efficient cavity-less optomechanical transduction involving opto-piezoelectric backaction from the bound photoexcited electron-hole pairs has been demonstrated in these systems, including self-feedback cooling, amplification of the thermomechanical motion, and control of the mechanical quality factor through carrier excitation.~\cite{17,18,19} The change in the electronic landscape produced by photoexcited carriers also induces a stress in the structure through electron-phonon coupling mediated by deformation potential interaction. This stress can be identified as an optoelectronic force, and should have the same kind of temporal behavior (with different time-scales and details depending on the carrier dynamics) and amplified strength as observed for photothermal forces.~\cite{MetzgerPRL2008,MetzgerPRB2008,Restrepo} Recently optical cooling of mechanical modes of a GaAs nanomembrane forming part of an optical cavity was reported,~\cite{Usami} and its relation to optoelectronic stress via the deformation potential was analysed. Because of the very fast relaxation rate of excited carriers due to surface recombination in such nanometer size structures, it was concluded that thermal (and not optoelectronic) stress was the primary cause of cooling in that case. We will demonstrate here that the carrier dynamics can be fundamentally modified when the nanometer size GaAs layer is embedded in a monolithic microcavity, making optoelectronic forces the  main mechanism of interaction of light with vibrations in such semiconductor devices. The diffusion of the photoexcited carriers thus assumes a central role, a dynamics that can be engineered using embedded quantum wells (QWs).

Because of their optoelectronic properties, semiconductor GaAs/AlAs microcavities are interesting candidates for novel functionalities in cavity optomechanics. Perfect photon-phonon overlap, and access to electronically resonant coupling in addition to radiation pressure could lead to strong optomechanical interactions of photoelastic origin.~\cite{FainsteinPRL2013,Baker,JusserandPRL2015} The vibrational frequencies in these microresonators are determined by the vertical layering of the device (fabricated with the ultra-high quality of molecular beam epitaxy), and not by the lateral pattering (defined by the more limited performance of microfabrication techniques). This has allowed access to much higher frequencies for the optomechanical vibrational modes, in the 20-100~GHz range, without significant reduction of the mechanical Q-factors.~\cite{AnguianoPRL} In addition, these optomechanical resonators enable the conception of hybrid architectures involving artificial atoms (semiconductor excitons) coupled to the optical cavity mode, and thus combining the physics of cavity optomechanics with cavity quantum electrodynamics.~\cite{Rozas_Polariton,Restrepo2,Kyriienko}  Our motivation here is to search for optoelectronic forces in these devices, and for that purpose we study the light-sound coupling involving the resonator mechanical modes and the optical cavity at resonance with the material exciton transition energy. Clear evidence of the role of optoelectronic forces emerges from new studies based on the spectral dependence and the relative intensity of the observed mechanical cavity modes. We demonstrate based on these experiments and model calculations that the main phonon generation mechanism using pulsed lasers close to resonance in these devices involves indeed the real excitation of carriers and the deformation potential mechanism.~\cite{Baker,RuelloUltrasonics} We show that in microcavities with ``bulk'' GaAs spacers (i.e. cavity spacers constituted by a thick $\lambda/2$ layer of GaAs) ultrafast carrier redistribution leads to an enhanced coupling to the more uniformly distributed fundamental 20~GHz cavity vibrational mode. The relative intensity of the modes in these structures varies with laser power, consistently with a more uniform distribution of carrier being attained at higher concentrations. An engineering of the structure taking into account this effect and using embedded quantum wells is used to limit the carrier diffusion, leading to mechanical modes with a relative intensity consistent with the spatial distribution of the cavity optical field.  The demonstration of optoelectronic forces and the possibility to tune the coupling to specific vibrations using quantum wells opens the way to new opportunities in the field of optomechanics.

\section{Results}

\begin{figure}[!t]
    \begin{center}
    \includegraphics[trim = 0mm 0mm 0mm 20mm,clip,scale=0.45,angle=0]{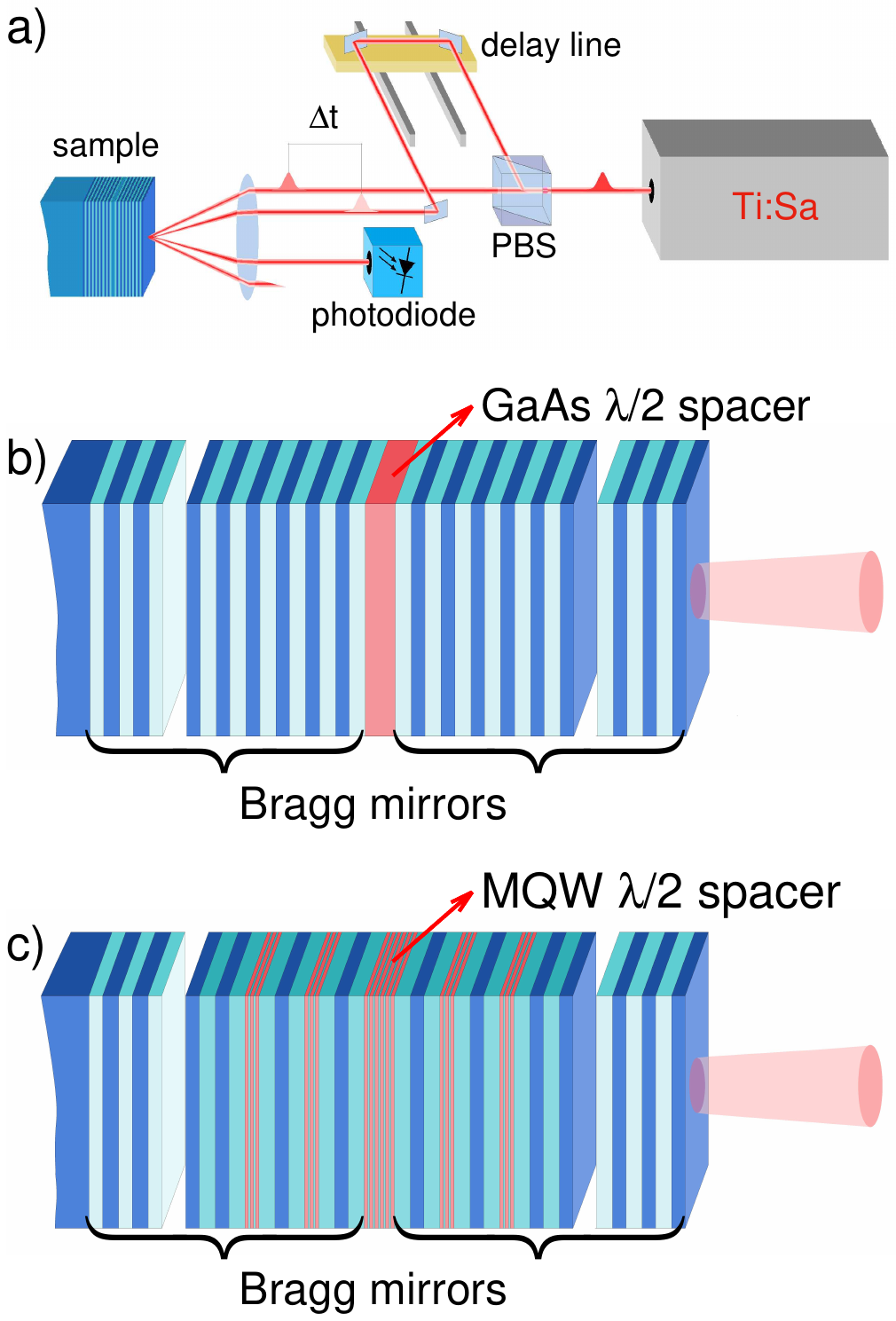}
    \end{center}
    \vspace{-0.8 cm}
\caption{(Color online) (a) Time resolved reflectance difference experimental set-up. PBS stands for polarizing beam splitter. (b) and (c) are schemes of the GaAs ``bulk'' and ``MQW'' microcavities, respectively.\label{Fig1}}
\end{figure}

We consider two {\em planar} microcavity structures, specifically a ``bulk'' GaAs and a multiple quantum well (MQW) resonator. The ``bulk'' GaAs microcavity is made of a $\lambda/2$ GaAs-spacer enclosed by ($\lambda/4,\lambda/4)$ Al$_{0.18}$Ga$_{0.82}$As /AlAs  DBRs, 20 pairs of layers on the bottom, 18 on top, grown on a GaAs substrate (a scheme of the structure is  presented in Fig.~\ref{Fig1}(b)).~\cite{Tredicucci,AFainsteinBulkGaAs}
%The DBRs stop band is centered at the design wavelength of $817~\mathrm{nm}$.
As we have demonstrated previously, this structure works as an optomechanical resonator that simultaneously confines photons and acoustic phonons of the same wavelength.~\cite{FainsteinPRL2013,Trigo,Anguiano,PSesin,AnguianoPRL} In the MQW microcavity the $\lambda/2$ spacer is constituted by six $14.5~\mathrm{nm}$ GaAs QWs separated by $6.1~\mathrm{nm}$ AlAs barriers. To further enhance the light-sound coupling the second and fourth $\lambda/2$ DBR alloy layers on each side are also replaced by three GaAs/AlAs QWs. The reason for this design will become clear below. The DBRs in this case are ($\lambda/4,\lambda/4)$ Al$_{0.10}$Ga$_{0.90}$As /AlAs  multilayers, 27 pairs on the bottom, 23 on top, grown again on a GaAs substrate. A scheme of this structure is displayed in Fig.~\ref{Fig1}(c). The number of DBR periods in both structures is designed to assure an optical Q-factor $Q \geq 10^4$ (cavity photon lifetime $\tau \sim 5$~ps). The samples have a thickness gradient so that the energy of the optical cavity mode, and its detuning respect to the bulk GaAs and MQW gaps, can be varied by displacing the laser spot on the surface.

Reflection-type degenerate pump-probe experiments were performed with the laser wavelength tuned with the optical cavity mode (see the scheme in Fig.~\ref{Fig1}(a)).~\cite{Thomsen,Bartels}
Picosecond pulses ($\sim 1$~ps) from a mode-locked Ti:Sapphire laser, with repetition rate 80~MHz, were split into cross polarized pump (power 20~mW) and probe (1~mW) pulses.  Both pulses were focused onto superimposed $\sim 50$ $\mu$m-diameter spots. To couple the light to the microcavity the probe beam propagates close to  the sample normal and is tuned to the high derivative flank of the cavity mode reflectivity dip, while the pump incidence angle is set for resonant condition precisely at the cavity mode.~\cite{Kimura_coherentcavity,KimuraTheory,Kimura_doubleresonance} The laser wavelength was also set so that the phonon generation and detection would be close to resonance with the direct bandgap of the GaAs ``bulk'' cavity spacer ($E_\textrm{\tiny GaAs} \sim 1.425$~eV) or of the QWs ($E_\textrm{\tiny QW} \sim 1.526$~eV). To accomplish this resonant excitation the temperature was also used as a tuning parameter; the ``bulk'' cavity experiments were done at room temperature, while the MQW structure was studied at 80~K.  Light is thus coupled resonantly with the optical cavity mode and the semiconductor excitonic resonance. Acoustic phonons confined in the same space as the optical cavity mode are selectively generated within the resonator, and are detected through their modulation of the optical cavity mode frequency.\cite{FainsteinPRL2013}

These pump-probe ultrafast laser experiments are conceptually similar to the ring-down techniques recently exploited in the cavity optomechanics domain,~\cite{Usami} but more appropriate to the study of ultra-high frequency vibrations (GHz-THz range.) The pulsed laser phonon generation efficiency can be described as:~\cite{Flor2012}
\begin{equation}
g(\omega)\propto \int{\kappa(z)\eta_0(\omega, z)|F_p(z)|^2}dz.
\label{eq2}
\end{equation}
Here $\omega$ is the phonon frequency, $\eta_0$ describes the elastic strain eigenstates, $F_p(z)$ is the spatially dependent perturbation induced by the pump laser,  and $\kappa$ is an effective material-dependent generation parameter that considers different light-matter couplings. All parameters are implicitly assumed to depend on the laser wavelength. We will be interested here in the relative intensity of the vibrational modes, not in their absolute values, so the main physical ingredients are expressed in the functional form of Eq.~\eqref{eq2}. This equation reflects the spatial overlap of the strain eigenstates with the light-induced stress. The latter can be written (independently of the mechanism involved) as $\sigma_p(z,t)=\kappa(z)|F_p(z)|^2T(t)$.~\cite{Flor2012} Here $T(t)$ is the function describing the temporal evolution of the light-induced perturbation. Typically it is a delta-like function for radiation pressure and electrostriction forces, and a step-like function for the photothermal and optoelectronic mechanisms, broadened by the time-delay of the mechanism involved. The spatial distribution of the optical excitation $F_p(z)$ along the growth axis ($z$) corresponds to the cavity confined electric field $E_c(z)$ for radiation pressure and electrostriction forces, but can be different from it for the other two mechanisms depending on the spatial distribution and dynamics of excited charges and laser-induced temperature variations. As we argue next, this will be a way to identify the main optical force under play in the studied devices.

\begin{figure*}[!t]
    \begin{center}
    \includegraphics[trim = 0mm 0mm 0mm 10mm,clip,scale=0.6,angle=0]{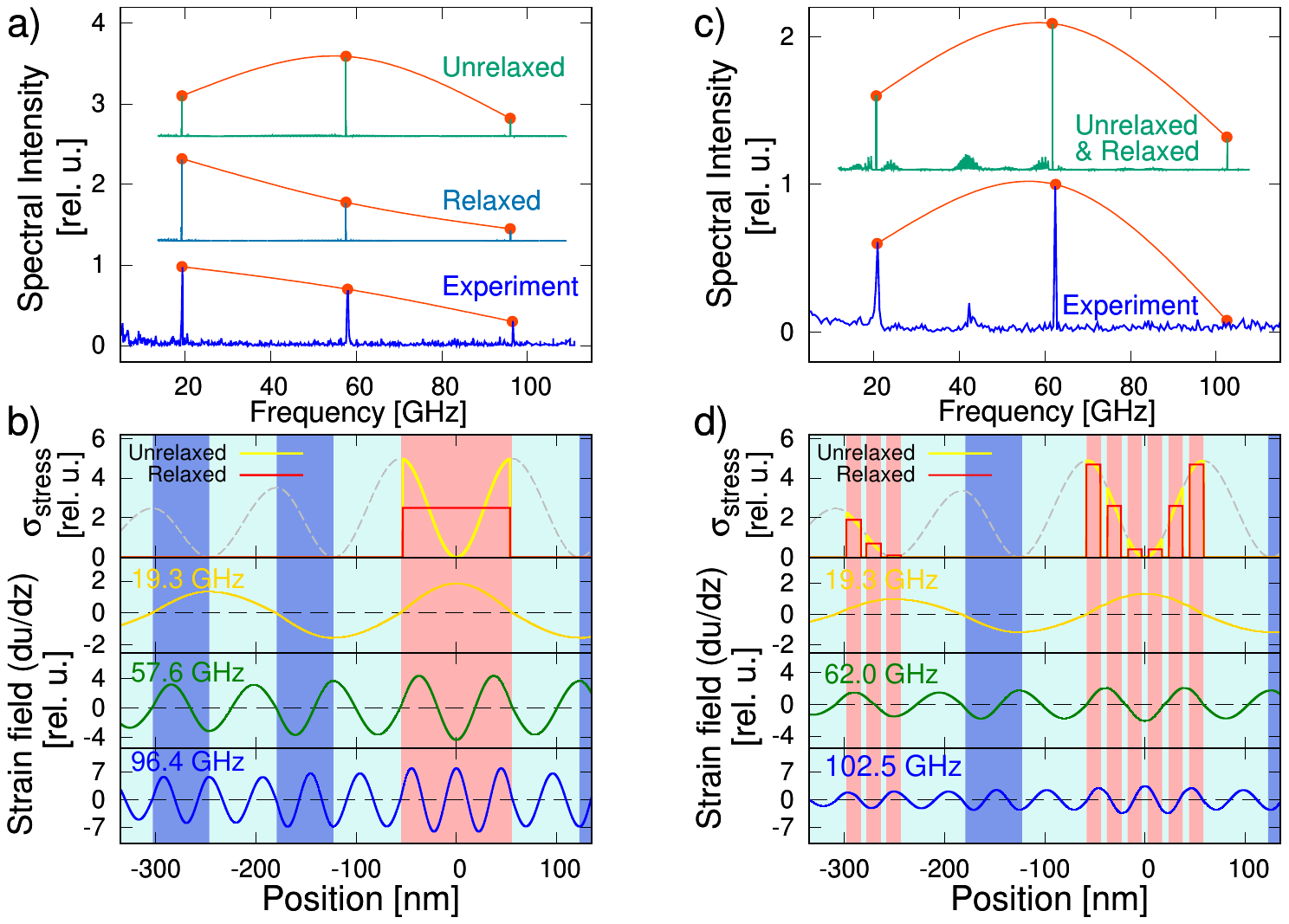}
    \end{center}
    \vspace{-0.8 cm}
\caption{(Color online) (a) and (c): Measured and calculated acoustic phonon spectra for excitation resonantly tuned to the optical cavity mode for the GaAs and QW cavities, respectively. The laser energy was set around 10~meV below the gap of either bulk GaAs, or the MQW exciton transition. The calculations consider a resonant generation only at the GaAs layers, either with a spatial pattern along the growth direction reproducing the squared modulus of the pump field (unrelaxed), or a flat generation that assumes a rapid distribution of carriers within the full width of the corresponding GaAs layer (relaxed). (b) and (d) present the excitation pattern for the two considered situations, and strain distribution associated to the first three confined acoustic modes corresponding to the GaAs and MQW cavity structures, respectively. The position $z=0$ marks the center of the cavity spacer. Solid yellow curves in the top panels correspond to the unrelaxed spatial pattern of the optical stress. The red step-like solid lines indicate the stress for the relaxed situation. The grey dashed curve shows how the cavity confined field is distributed through the cavity structure, but there is no optical force in the dashed regions because there are no excited photocarriers due to the much larger gaps of the involved materials.\label{Fig2}}
\end{figure*}

Figures~\ref{Fig2} a-b present the case of the ``bulk'' GaAs cavity. For the experiments reported here the laser was set around 10~meV below the energy of the bulk GaAs gap. Panel (a) in the figure displays the experimental spectrum, compared with calculations assuming that $F_p(z)$ reproduces the instantaneous spatial distribution of the light intensity  $|E_c(z)|^2$ (``unrelaxed''), or that it corresponds to a photoexcited carrier distribution within the GaAs-spacer that is flat along the growth direction (``relaxed''). That is, it assumes that within the time in which the pump laser-induced perturbation is effective (of the order of the half period of the vibrational frequencies involved), the spatial distribution of the photoexcited carriers relaxes extending their presence throughout the full width of the GaAs cavity spacer material.  Three peaks are clearly visible at $\sim 20$, $\sim 60$ and $\sim 100$~GHz, corresponding to the fundamental, second and fourth overtones of the z-polarized cavity confined breathing mechanical modes.~\cite{FainsteinPRL2013,AnguianoPRL} The associated spatial distribution of the strain fields $\eta_0(\omega,z)$, together with that of the light-induced optoelectronic stress, are shown in panel (b) of the figure. The solid yellow curves in the top panel corresponds to the unrelaxed spatial pattern of the optical stress. The red step-like solid line indicates the stress for the relaxed situation. The grey dashed curve shows how the cavity confined field is distributed, but there is no optical force in the dashed regions because photoelastic (electrostrictive) coupling is resonantly enhanced in GaAs, and photons are only absorbed in GaAs, all other materials being fully transparent at the involved wavelengths. That is, $\kappa(z)$ is assumed to be non-zero {\em only} in GaAs. The solid curves in $\sigma_\textrm{\tiny stress}$ thus represent the region where the light-induced stress is finite, either reflecting the excitation field (yellow), or the relaxed situation (red). It is clear in the experiment that the vibrational mode's amplitude decreases systematically with increasing frequency of the mode, something that according to the calculations is only compatible with the carriers having rapidly spread filling the full width of the cavity spacer along the growth direction. The explanation is straightforward considering the overlap integral given by Eq.~\eqref{eq2}, and the involved spatial distributions depicted in panel (b) of Fig.~\ref{Fig2}. It clearly excludes radiation pressure and electrostriction as the possible driving mechanisms.

Based on the above discussion it is also clear from the top curve in Fig.~\ref{Fig2}(a) that the relative weight of the higher frequency 60~GHz mode could be enhanced if the spatial distribution of the optical perturbation $F_p(z)$ could be forced to map-out the cavity field intensity $|E_c(z)|^2$. If, as argued, the main generation mechanism is indeed governed by optoelectronic forces, one could accomplish this task by artificially limiting the diffusion of the photoexcited carriers along the growth axis. A natural way to do this is through an adequate  engineering of the cavity spacer, e.g. by introducing GaAs/AlAs MQWs.  This case is shown in Figs.~\ref{Fig2}(c-d). Again panel (c) shows the experiment and calculation of the coherent phonon spectral intensity, with the laser energy set this time around 10~meV below the MQW exciton transition energy. Panel (d) displays the spatial distribution of the photoexcited stress, and that of the strain related to the involved vibrational modes. In this case relaxing the carrier distribution to fill the full width of the QWs (red step-like curves), or maintaining the exact laser excitation pattern (yellow solid curve), makes no observable difference in the calculated spectra. Interestingly, by tailoring the spatial distribution of the photoexcited carriers using quantum wells we are able to confirm the role of optoelectronic stress as the main optomechanical coupling, and furthermore we have pushed the main vibrational frequency of these optomechanical resonators from the fundamental mode at 20~GHz to the second overtone at 60~GHz. These frequencies are one order of magnitude larger than the record frequencies demonstrated in other cavity optomechanics approaches.~\cite{ReviewCOM} From Fig.~\ref{Fig2}(d) it also becomes clear why additional QWs were introduced in the design at the second and fourth DBR periods, and {\em not} at the first and third: because of a change of sign of the strain fields, the latter contribute to the overlap integral in Eq.~\eqref{eq2} with the sign reversed respect to the cavity spacer.

\begin{figure}[!t]
    \begin{center}
    \includegraphics[trim = 0mm 0mm 0mm 0mm,clip,scale=0.35,angle=0]{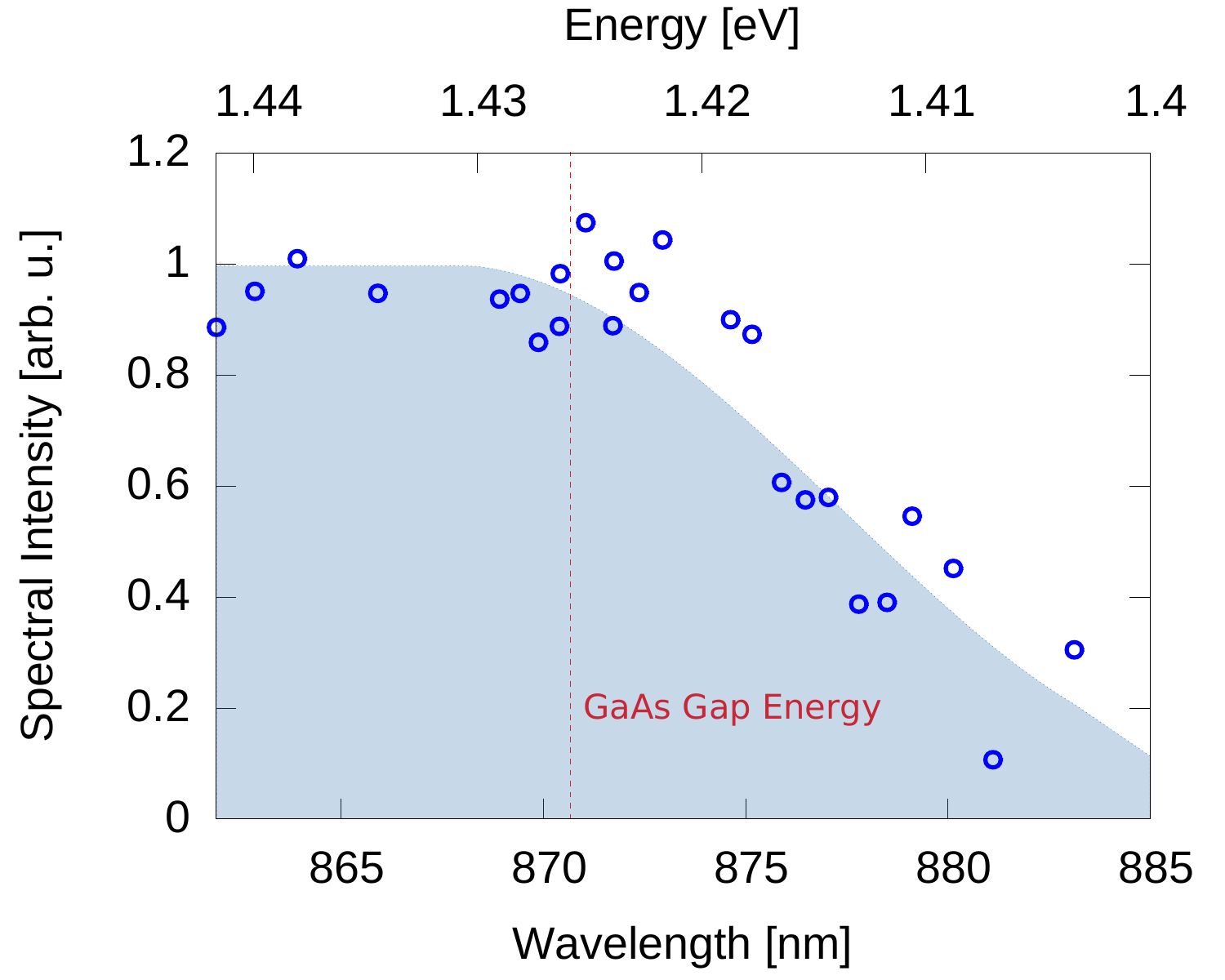}
    \end{center}
    \vspace{-0.8 cm}
\caption{(Color online) Wavelength dependence of the spectral intensity corresponding to the 60~GHz mode in the bulk-GaAs cavity . The vertical line indicates the energy of the GaAs gap. \label{Fig3}}
\end{figure}

The role of photoexcited carriers in the optical forces involved in the described coherent phonon generation with pulsed lasers can be furthermore verified by investigating the laser wavelength dependence of the mechanical mode intensity. This is shown for the 60~GHz mode of the bulk-GaAs cavity in Fig.~\ref{Fig3}. Note that each point in this figure corresponds to an experiment in which the laser energy is varied and the position of the spot in the tapered structure is accordingly changed so that the tuning with the optical cavity mode remains the same. The wavelength limits of the experiment are thus defined by the thickness gradient existent in the microcavity structure.  A threshold-like behaviour with spectral intensity tending to zero in the transparency region of the structure, and finite intensity only close and above the gap of GaAs, is observed. This is indicative of real electron-hole pairs being the mediators of the involved optical force.

\begin{figure}[!t]
    \begin{center}
    \includegraphics[trim = 0mm 0mm 0mm 0mm,clip,scale=0.29,angle=0]{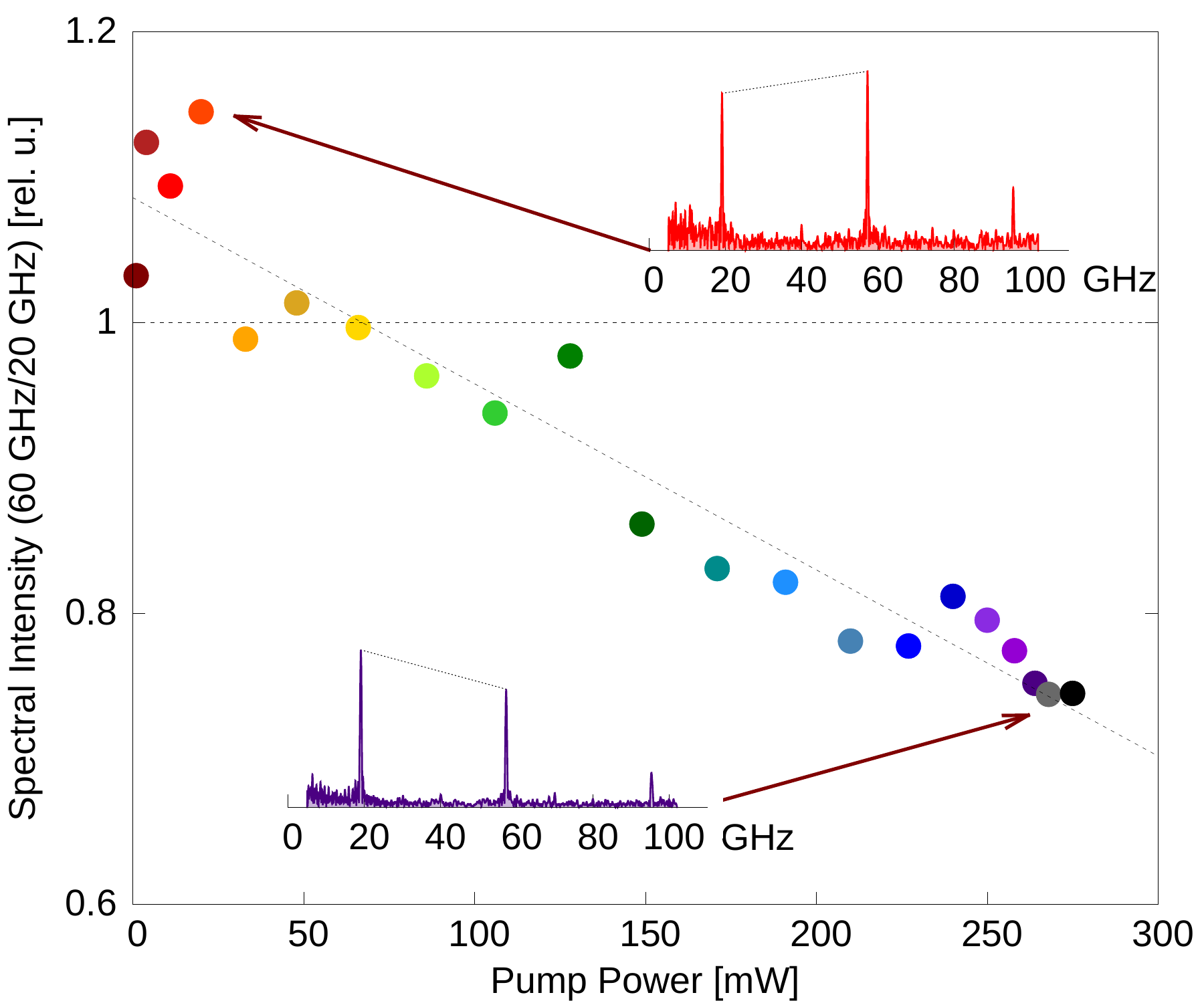}
    \end{center}
    \vspace{-0.8 cm}
\caption{(Color online) Relative intensity of the 60 and 20~GHz modes as a function of pump laser power for the bulk-GaAs microcavity obtained with the laser tuned around 10~meV below the gap. Characteristic spectra at the two extreme limits of studied laser power are shown as insets. Note the progressive change of the relative intensity of the modes as the laser power increases. \label{Fig4}}
\end{figure}

Note that although the spectrum calculated for the relaxed situation of the bulk-GaAs microcavity in Fig.~\ref{Fig2}a shows the same tendency as the experimental result, a quantitative difference between experiment and theory still remains. Namely, the relative intensity of the 60~GHz mode is slightly larger than predicted. The most natural explanation for this remaining difference between the bulk-GaAs cavity experiments and the relaxed calculation is that relaxation throughout the whole thick GaAs spacer of the cavity is not complete within the relevant times involved in the coherent generation process (times typically of the order of half a period, i.e., around 7~ps for the 60~GHz mode). Immediately after excitation, the optical force has to reproduce the spatial pattern of the cavity confined optical field (such force leads to larger intensity for the 60~GHz mode). This pattern rapidly relaxes towards a uniform distribution along the growth axis within the GaAs material (a distribution in the optical force that provides larger intensity for the 20~GHz mode, as shown in Fig.\ref{Fig2}). If this relaxation is fast in comparison with the mechanical period, the latter will apply. If the relaxation is not complete, one can expect a behaviour in between the two limits as observed experimentally. One way to test this hypothesis is to perform experiments for a varying concentration of photoexcited carriers, as shown in Fig.~\ref{Fig4}, where the relative intensity of the 60 and 20~GHz modes as a function of pump laser power is displayed. One can expect that due to electron-electron Coulomb interactions higher carrier densities tend to accelerate the homogenisation of carriers within their available space. This is indeed what is evidenced in the experiments, with a progressive trend towards the relaxed situation, i.e. a flat distribution of the optical stress consistent with the 20~GHz mode being stronger than the 60~GHz one, as the pump laser power is increased.

Having demonstrated the central role of optoelectronic forces in the generation of GHz phonons in semiconductor microcavities upon resonant pulsed excitation,  we address next the potential of this mechanism for the observation of laser cooling and optically induced self-oscillation. The fundamental concepts describing backaction dynamics in cavity optomechanics are grasped by the delayed force model which, for the optically modified vibrational damping rate $\Gamma_{\mathrm{eff}}$, gives~\cite{MetzgerPRB2008}
\begin{equation}
\Gamma_\textrm{\tiny eff}=\Gamma_\textrm{\tiny M}\left( 1+Q_\textrm{\tiny M}\frac{\Omega_\textrm{\tiny M} \tau}{1+\Omega^2_\textrm{\tiny M} \tau^2}\frac{\nabla{F}}{K} \right),
\label{eq1}
\end{equation}
with $Q_\textrm{\tiny \tiny M}$ and $\Gamma_\textrm{\tiny M}=\Omega_\textrm{\tiny M}/Q_\textrm{\tiny M}$  the unperturbed mechanical quality factor and damping, respectively. $\Omega_\textrm{\tiny M}$ and $K$ are respectively the unperturbed mechanical mode frequency
and stiffness, while $\nabla{F} = \partial F/\partial u|_{u_0}$ represents the change in the steady-state optical force for a small displacement $\delta u$ of the mechanical resonator around the equilibrium position. To lead to backaction optical forces need to respond with a delay to fluctuations that change the frequency of the optical mode. The simplest delay function to consider is an exponential function $h(t) = 1 - \textrm{exp} (t/\tau)$, with $\tau$ the corresponding time-delay. Typically $\tau$ would correspond to the cavity photon lifetime $\tau_\textrm{\tiny cav}$, but in general and particularly for photothermal and optoelectronic forces, it can be significantly longer than $\tau_\textrm{\tiny cav}$. $\Gamma_\textrm{\tiny eff}$ in the presence of dynamical backaction can thus increase if $\nabla{F} > 0$, leading to laser cooling, or decrease and eventually attain zero (self-oscillation) if $\nabla{F} < 0$. The magnitude of this effect is proportional to the gradient of the optical force, which is a function of the deposited optical energy and the involved optomechanical coupling mechanism (that is, how this deposited energy is translated into a mechanical deformation). It is also proportional to a delay tuning factor $f_D = \frac{\Omega_\textrm{\tiny M} \tau}{1+\Omega^2_\textrm{\tiny M} \tau^2}$. $f_D$  has a maximum for $\Omega_\textrm{\tiny M} \tau = 1$, which reflects the intuitive fact that the force fluctuations are more effective to induce vibrations if their time-constant is neither too short, nor too long, but tuned to the vibrational frequency so that $\tau \approx 1/\Omega_\textrm{\tiny M}$.

\begin{table}[t]
\centering
\begin{tabular}{|c|c|c|c|}
\hline
Mechanism & Opt. force (N) & $\textrm{ f}_{D} =\frac{\Omega_m \tau}{1+(\Omega_m \tau)^2}$  \\
\hline
Rad. Pressure &  $2.8 \times 10^{-14}$ & $\sim 0.44$ \\
\hline
Electrostriction &  $7.6 \times 10^{-14}-8.1\times10^{-13}$ & $\sim0.44$ \\
\hline
Photothermal &  $1.0 \times 10^{-11}$  & $\sim 10^{-5}$ \\
\hline
Optoelectronic &  $3.7 \times 10^{-11}$ & $\sim 0.04$ \\
 \hline
\end{tabular}
\caption{\label{table} Optical force (modulus) per trapped (or absorbed) photon, and the value of the delay tuning factor $f_D$ for the different optical forces present in the GaAs microcavity. The forces have been evaluated using finite-element methods and considering a cylindrical pillar of 2 microns diameter. For the case of electrostriction two values are given, corresponding to room temperature out of resonance ($1.15 \mu$m) and slightly below resonance (870~nm) values of the photoelastic constant.~\cite{JusserandPRL2015}}
\end{table}
Table~\ref{table} presents the different factors intervening in Eq.~\eqref{eq1} for the studied DBR microcavities and the considered mechanisms, namely, the corresponding magnitude of the involved optical forces and the delay tuning factor $f_D$. The optical forces are given per photon (trapped in the cavity or absorbed depending on the mechanism). They have been evaluated using finite-element methods (see Appendix \ref{sec: appendix}),~\cite{Baker,MetzgerPRB2008} and considering a micro-pillar with 20 period DBRs of 2 micrometer diameter. As argued above, the main difference between photothermal and optoelectronic forces respect to radiation pressure and electrostriction, is the larger amount of deposited energy per trapped cavity photon. Similarly to what is observed in GaAs microdisks,~\cite{Baker} well below the gap (far from the excitonic resonances) both geometric (radiation-pressure) and photoelastic contributions to the optomechanical coupling factor have similar values.  Ultra-strong resonant enhancement of the photoelastic coupling has been experimentally demonstrated in bare GaAs/AlAs MQWs.~\cite{JusserandPRL2015} In Table~\ref{table} we provide the magnitude of the electrostrictive force per incident photon considering the two situations, far from resonance and at resonance as used in our experiments, assuming that the room temperature values of the photoelastic constant given in Ref.\,\onlinecite{JusserandPRL2015} are valid for a similar MQW embedded in a pillar microcavity.   Concerning the photothermal coupling, a quantitative evaluation of its relevance in semiconductor GaAs membranes and microdisks indicates that it can be significant.~\cite{Baker,Usami}  It is also significant in our microcavities, as evidenced in Table~\ref{table}.  What excludes it as a potentially relevant optomechanical force is its slow dynamics, which for the high frequency vibrations considered here leads to  delay tuning factors $f_D \sim 10^{-5}$ even for the fundamental breathing mode at 20~GHz and considering a relatively fast thermal relaxation $\tau$ of the order of a $\mu$s. Note that due to the deformation potential interaction in GaAs, optoelectronic forces have the same sign as the thermal stress and contribute to expand the crystal (they have the reverse sign in Si).~\cite{RuelloUltrasonics} The delay tuning factor is close to its maximum value 0.5 for the impulsive radiation pressure and electrostrictive mechanisms, considering a cavity photon lifetime of a few picoseconds as in our vertical microcavities. It is also reasonably well tuned for the optoelectronic forces and 200~ps recombination times as demonstrated for pillars of a few microns lateral size in Ref.\,\onlinecite{AnguianoPRL}. In fact, $f_D \sim 0.44$ is precisely the same value as attained for optimized cantilever resonators that have evidenced strong optical cooling and self-oscillation dynamics based on photothermal forces.~\cite{MetzgerPRL2008,MetzgerPRB2008,Restrepo} The factor $1/10$ decrease of $f_D$ with respect to radiation pressure and electrostriction is overly compensated by the larger efficiency of optoelectronic forces. Note that the physical mechanism at the base of optoelectronic forces is the same as in photoelastic coupling, namely, deformation potential interaction. The qualitative difference is that a photon is scattered in the latter, while it is  absorbed with the subsequent creation of real electron-hole pairs in the former.

\section{Conclusions and Outlook}

In conclusion, we have shown that optoelectronic deformation potential interactions are at the origin of the optical forces acting for excitation with pulsed lasers close to the semiconductor gaps in GaAs/AlAs microcavities. The carrier dynamics following photoexcitation is determinant in the emitted coherent acoustic phonon spectrum, and can be tailored using quantum wells to push the vibrational optomechanic frequencies from 20 up to 60~GHz, an order of magnitude larger than the highest standards in cavity optomechanics.  The strong potential of optoelectronical forces for the demonstration of backaction dynamical effects in semiconductor microcavities was addressed. This could open the way to ultra-high frequency cavity optomechanics, and through it to quantum measurements and applications at higher temperatures than currently accessible.

\section*{Funding Information}
This work was partially supported by the ANPCyT Grants PICT 2012-1661 and 2013-2047, the Labex NanoSaclay,  and the international franco-argentinean laboratory LIFAN (CNRS-CONICET).

\appendix
\section{Details regarding the calculation of the optical forces.}\label{sec: appendix}

We describe the optomechanical coupling of a fundamental optical cavity mode with the fundamental acoustic cavity mode.\cite{FainsteinPRL2013} For a normalized displacement mode $\vec{u}(\vec{r})$, we can parametrize the profile as $\vec{U}(\vec{r}) = u_0 \, \vec{u}(\vec{r})$. The effective mass is obtained by the requirement that the potential energy of this parametrized oscillator is equal to the actual potential energy:
\begin{equation}
\label{pot}
\frac{1}{2}\Omega_{\text{M}}^2 \int d\vec{r} \rho(\vec{r}) |\vec{U}(\vec{r})|^2 = \frac{1}{2} m_\text{eff} \Omega_\text{M}^2 u_0^2.
\end{equation}
The effective mass $m_\text{eff}$ is,
\begin{equation}
\label{m_eff}
m_\text{eff} = \frac{\int d\vec{r} \rho(\vec{r}) |\vec{U}(\vec{r} )|^2}{u_0^2} \equiv \int d \vec{r} \rho(\vec{r}) |\vec{u}(\vec{r})|^2,
\end{equation}
where $\rho(\vec{r})$ is the scalar density distribution field for the structure.
As discussed in Ref.\,\onlinecite{Baker}, we consider the normalization for the mechanical modes such that the position $\vec{r}_0$ (known as the reduction point and chosen so that the displacement is maximum) satisfies $| \vec{u}(\vec{r}_0) |= 1$. In our system, the reduction point lies at the interfaces of the cavity spacer of GaAs.
The equation of motion for the cavity breathing mode is modeled as an oscillator described by the displacement $u$, with an effective mass m$_\text{eff}$, mechanical damping $\Gamma_\text{M}$, and stiffness constant $\text{K}=\text{m}_\text{eff}\Omega^2_\text{M}$, given by\cite{MetzgerPRB2008}

\begin{equation}
\label{1Dproblem}
m_{\text{eff}} \frac{d^2u}{dt^2} + m_{\text{eff}} \Gamma_{\text{M}} \frac{du}{dt} +  m_{\text{eff}} \Omega_{\text{M}}^2 u = F_{geo} + F_{ph} + F_{th} + F_{oe}.
\end{equation}

The right-hand side corresponds to a sum over all the optical forces that drive the mechanical system: geometrical related to radiation pressure ($F_{geo}$),\cite{Baker} photoelastic ($F_{ph}$), thermoelastic ($F_{th}$) and optoelectronic ($F_{oe}$).

We proceed to the evaluation of the forces. For the electrostrictive and geometrical forces we can obtain the corresponding values computing $F = \hbar g_{om}$, where $g_{om}$ is the geometrical or photoelastic optomechanical coupling constant.\cite{Baker}

For the calculation of the geometric optomechanical coupling factor $g_{om}^{geo}$, we follow the analysis proposed by Johnson \textit{et al},\cite{Johnson2002} implementing a finite element-method to obtain the electric and acoustic fields. We generalize the approach presented in Refs.\,\onlinecite{Ding2010,Baker} to compute the effects induced by the multiple interfaces at the DBR's boundaries,
\begin{equation}
\label{geometric_one}
g_{om}^{geo} =  \frac{\omega_c}{2} \sum_i \frac{\oint_{A_i}  (\vec{u} \cdot \hat{n_i}) (\Delta \epsilon_i|\vec{E}_{\|}|^2- \Delta(\epsilon_i^{-1}) |\vec{D}_{\perp}|^2 ) dA_i}{\int \epsilon |\vec{E}|^2 d\vec{r}},
\end{equation}
where $\omega_c$ is the optical angular frequency at resonance, $\vec{u}$ the normalized displacement field, $\hat{n}_i$ the unitary normal-surface vector corresponding to the interface, $\Delta \epsilon_i=\epsilon_{i,left} - \epsilon_{i,right}$ the difference between the dielectric constants of the materials involved, $\Delta \epsilon_i^{-1}=\epsilon_{i,left}^{-1} - \epsilon_{i,right}^{-1}$, $\vec{E}_{\|}$ is the component of the electric-field parallel to the interface surface and $\vec{D}_{\perp}$ is the normal component of the displacement field $\vec{D} = \epsilon_0 \epsilon_r \vec{E}$. The index $i$ runs over every distinct interface $A_i$.

The photoelastic contribution to the optomechanical coupling occurs due to the strain-field modulation of the dielectric properties, i.e. $\Delta (\frac{1}{\epsilon_r})_{ij} = p_{ijkl} S_{kl}$, and is given by,\cite{Baker}
\begin{equation}
\label{eq:gc}
g_{om}^{ph} = \frac{\omega_c  \epsilon_0 }{2} \frac{\int  n^4  E_i p_{ijkl} S_{kl} E_j d\vec{r}}{\int \epsilon |\vec{E}|^2 d\vec{r}},
\end{equation}
where $\epsilon=\epsilon_0 \epsilon_r(\vec{r})$ is the dielectric function. Due to the resonant character\cite{JusserandPRL2015} we consider $p_{ijkl}$ to be non-vanishing only in the GaAs spacer. Only three different components for this tensor are non-zero due to the cubic symmetry. Since the non-diagonal component of the strain $S_{rz}$ is non-zero we also take $p_{44} = (p_{11}-p_{12})/2$ into account.\cite{Florez2016} Through Raman experiments, B. Jusserand \textit{et al} have determined the wavelength and temperature dependence for the GaAs photoelastic constant $p_{12}$.\cite{JusserandPRL2015} On the contrary, there is no similar reported information for $p_{11}$. However, because of the prevalent z-polarized character of the acoustic modes, it turns out that the contribution to $g^{ph}_{om}$ is dominated by $p_{12}$, which reaches $p_{12} = 1.526 $ for $1.42\,$eV at room temperature.\cite{JusserandPRL2015}

For thermoelastic and optoelectronic effects we used the expression

\begin{equation}
\label{Forces}
F(t) = \int_V \frac{\sigma_{ij}(r,t) S_{ij}(r,t)}{u(t)} d\vec{r},
\end{equation}
where $S_{ij}$ is the strain field tensor related to the mechanical breathing mode $u$, and $\sigma_{ij}$ is the stress field tensor related to the driving mechanism. The stress tensor for the thermoelasticity can be determined by \cite{RuelloUltrasonics}
\begin{equation}
\label{thermo-stress}
\sigma_{th} = - \gamma_\text{L} C_\text{L} \Delta T_\text{L}(r,t),
\end{equation}
where $C_\text{L}$ is the heat capacity, $\gamma_\text{L}$ is the Gr\"uneisen coefficient \cite{Eryiit1996} and $\Delta T_\text{L}(r,t)$ the lattice temperature variation. Assuming a complete transfer of energy from the electronic to the phononic system, considering intraband and non-radiative interband relaxation processes for excited electron-hole pairs we can determine $\Delta T_\text{L}$\cite{RuelloUltrasonics}. For the intraband decay channel, the temperature variation can be computed as $\Delta T_\text{L} = N_e (\hbar \omega - E_\text{G})/C_\text{L}$, where $\hbar \omega$ is the driving energy and $N_e$ represents the photoexcited carriers population density. The non-radiative interband relaxation processes give a temperature variation that can be accounted as $\Delta T_\text{L} = N_e E_\text{G}/C_\text{L}$ when the excitation is resonant with the bandgap energy $E_\text{G}$.

For the optoelectronic contribution we consider the limiting case in which electron-hole pairs are excited resonantly with the bandgap energy, and the dominant term in the electronic self-energy due to optical excitation corresponds to the variation in $E_\text{G}$. This stress can be summarized as follows,\cite{RuelloUltrasonics}
\begin{equation}
\label{opto-stress}
\sigma_\text{oe} = - d_{eh} N_e,
\end{equation}
where $d_{eh}$ is the deformation potential coefficient ($\sim 9\,$eV for GaAs.\cite{RuelloUltrasonics})

For the geometric and photoelastic effects the forces are given per number of photons.\cite{Baker} In order to compare,  for the thermoelastic and optoelectronic cases the forces are given per number of excited electron-hole pairs. For this purpose $N_e$ is considered equal to  $1/V_\text{eff}$ (the inverse of the GaAs spacer volume where carriers are optically excited).

In Table \ref{tab1} we present relevant optomechanical parameters and the magnitude of the calculated forces involved. We conclude that the optoelectronic forces have similar magnitude as the thermal forces and are several orders of magnitude greater than the photoelastic and radiation-pressure mechanisms ($\sim 10^2-10^3$).

\begin{table}
\begin{tabular}{c c}
Property & Bulk GaAs Cavity\\
\hline
$m_\text{eff}$ & 0.52 pg \\
$\Omega_\text{M}/2\pi$ & 19.66 GHz \\
$g_{om}^{geo}/2\pi$ & 42.4 GHz/nm \\
$g_{om}^{ph}/2\pi$ & 1.22 THz/nm \\
$g_{0}^{geo}/2\pi$ & 38.4 KHz \\
$g_{0}^{ph}/2\pi$ & 1.10 MHz \\
%$F_\text{geo}$ & 2.8 $10^{-14}$ N \\
%$F_\text{ph}$ & 8.1 $10^{-13}$ N \\
%$F_\text{th}$ & 1.0 $10^{-11}$ N \\
%$F_\text{oe}$ & 3.7 $10^{-11}$ N \\
\hline
\end{tabular}

\caption{ Optomechanical properties for bulk GaAs micropillar resonator.}
\label{tab1}
\end{table}

%\bigskip \noindent See \href{link}{Supplement 1} for supporting content.


\begin{thebibliography}{1}

\bibitem{ReviewCOM} M. Aspelmeyer, T. J. Kippenberg, and F. Marquardt, "Cavity optomechanics," Rev. Mod. Phys. {\bf 86}, 1391 (2014).

\bibitem{Ligo} B. P. Abbott et al., "GW150914: The Advanced LIGO Detectors in the Era of First Discoveries - LIGO Scientific and Virgo Collaborations," Phys. Rev. Lett. {\bf 116}, 131103 (2016).

% Optomechanical resonators in quantum limit


\bibitem{O'Connell} A. D. O'Connell, M. Hofheinz, M. Ansmann, Radoslaw C. Bialczak, M. Lenander, Erik Lucero, M. Neeley, D. Sank, H. Wang, M. Weides, J. Wenner,
John M. Martinis, and A. N. Cleland, "Quantum ground state and single-phonon control of a mechanical resonator," Nature {\bf 464}, 697 (2010).

\bibitem{Teufel} J. D. Teufel, T. Donner, Dale Li, J. H. Harlow, M. S. Allman, K. Cicak, A. J. Sirois, J. D. Whittaker, K. W. Lehnert, and R. W. Simmonds, "Sideband cooling of micromechanical motion to the quantum ground state," Nature {\bf 475}, 359 (2011).

\bibitem{Chan} J. Chan, T. P. Mayer Alegre, Amir H. Safavi-Naeini, Jeff T. Hill, Alex Krause, Simon Groeblacher, Markus Aspelmeyer, and Oskar Painter, "Laser cooling of a nanomechanical oscillator into its quantum ground state," Nature {\bf 478}, 89 (2011).

\bibitem{Verhagen} E. Verhagen, S. Deleglise, S. Weis, A. Schliesser, and T. J. Kippenberg, "Quantum-coherent coupling of a mechanical oscillator to an optical cavity mode," Nature {\bf 482}, 63 (2012).


% Optical Forces

\bibitem{Cohadon} P. F. Cohadon, A. Heidmann, and M. Pinard, "Cooling of a Mirror by Radiation Pressure," Phys. Rev. Lett. {\bf 83}, 3174 (1999).

\bibitem{Lin} Qiang Lin, Jessie Rosenberg, Xiaoshun Jiang, Kerry J. Vahala, and Oskar Painter, "Mechanical Oscillation and Cooling Actuated by the Optical Gradient Force," Phys. Rev. Lett. {\bf 103}, 103601 (2009).

\bibitem{Rakich1} P. T. Rakich, P. Davids, and Z. Wang, "Tailoring optical forces in waveguides through radiation pressure and electrostrictive forces," Opt. Express {\bf 18}, 14439-14453 (2010).

\bibitem{Rakich2} P. T. Rakich, C. Reinke, R. Camacho, P. Davids, and Z. Wang, "Giant Enhancement of Stimulated Brillouin Scattering in the Subwavelength Limit," Phys. Rev. X {\bf 2}, 011008 (2012).

\bibitem{FainsteinPRL2013} A. Fainstein, N. D. Lanzillotti-Kimura, B. Jusserand, and B. Perrin, "Strong Optical-Mechanical Coupling in a Vertical GaAs/AlAs Microcavity for Subterahertz Phonons and Near-Infrared Light,"
Phys. Rev. Lett. {\bf 110}, 037403 (2013).

\bibitem{Rozas_Polariton} G. Rozas, A. E. Bruchhausen, A. Fainstein, B. Jusserand, and A. Lema\^itre, "Polariton path to fully resonant dispersive coupling in optomechanical resonators," Phys. Rev. B {\bf 90}, 201302(R) (2014).

\bibitem{Baker} C. Baker, W. Hease, Dac-Trung Nguyen, A. Andronico, S. Ducci, G. Leo, and I. Favero, "Photoelastic coupling in gallium arsenide optomechanical disk resonators," Optics Express {\bf 22}, 14072 (2014).

\bibitem{MetzgerPRL2008} C. Metzger, M. Ludwig, C. Neuenhahn, A. Ortlieb, I. Favero, K. Karrai, and F. Marquard, "Self-Induced Oscillations in an Optomechanical System Driven by Bolometric Backaction," Phys. Rev. Lett. {\bf 101}, 133903 (2008).

\bibitem{MetzgerPRB2008} C. Metzger, I.Favero, A. Ortlieb, and K. Karrai, "Optical self cooling of a deformable Fabry-Perot cavity in the classical limit,"  Phys. Rev. B {\bf 78}, 035309 (2008).

\bibitem{Restrepo} J. Restrepo, J. Gabelli, C. Ciuti, and I. Favero, "Classical and quantum theory of photothermal cavity cooling of a mechanical oscillator," Comptes Rendus Physique
{\bf 12}, 860 (2011).

\bibitem{17} Hajime Okamoto, Daisuke Ito, Koji Onomitsu, Haruki Sanada, Hideki Gotoh, Tetsuomi Sogawa, and Hiroshi Yamaguchi, "Vibration Amplification, Damping, and Self-Oscillations in Micromechanical Resonators Induced by Optomechanical Coupling through Carrier Excitation," Phys. Rev. Lett. {\bf 106}, 036801 (2011)

\bibitem{18} Hajime Okamoto, Takayuki Watanabe, Ryuichi Ohta, Koji Onomitsu, Hideki Gotoh, Tetsuomi Sogawa, and Hiroshi Yamaguchi, "Cavity-less on-chip optomechanics using excitonic transitions in semiconductor heterostructures," Nature Comm. {\bf 6}, 8478 (2015).

\bibitem{19} Hajime Okamoto, Daisuke Ito, Koji Onomitsu, Tetsuomi Sogawa, and Hiroshi Yamaguchi, "Controlling Quality Factor in Micromechanical Resonators by Carrier Excitation," Applied Physics Express {\bf 2} (2009) 035001.

\bibitem{Usami} K. Usami, A. Naesby, T. Bagci, B. Melholt Nielsen, J. Liu, S. Stobbe, P. Lodahl, and E. S. Polzik, "Optical cavity cooling of mechanical modes of a semiconductor nanomembrane," Nature Physics {\bf 8}, 168 (2012).

% Polariton optomechanics

\bibitem{JusserandPRL2015} B. Jusserand, A.N. Poddubny, A.V. Poshakinskiy, A. Fainstein, and A. Lema\^itre, "Polariton Resonances for Ultrastrong Coupling Cavity Optomechanics in GaAs/AlAs Multiple Quantum Wells," Phys. Rev. Lett. {\bf 115}, 267402 (2015).

\bibitem{AnguianoPRL} S. Anguiano, A. E. Bruchhausen, B. Jusserand, I. Favero, F. R. Lamberti, L. Lanco, I. Sagnes, A. Lema\^itre, N. D. Lanzillotti-Kimura,  P. Senellart,
and A. Fainstein, "Micropillar Resonators for Optomechanics in the Extremely High 19-95--GHz Frequency Range," Phys. Rev. Lett. {\bf 118}, 263901 (2017).

\bibitem{Restrepo2} J. Restrepo, C. Ciuti, and I. Favero, "Single-Polariton Optomechanics," Phys. Rev. Lett. {\bf 112}, 013601 (2014).

\bibitem{Kyriienko} O. Kyriienko, T. C. H. Liew, and I. A. Shelykh, "Optomechanics with Cavity Polaritons: Dissipative Coupling and Unconventional Bistability," Phys. Rev. Lett. {\bf 112}, 076402 (2014).


\bibitem{RuelloUltrasonics} P. Ruello and V. E. Gusev, "Physical mechanisms of coherent acoustic phonons generation by ultrafast laser action," Ultrasonics {\bf 56}, 21 (2015).


% First report GaAs bulk

\bibitem{Tredicucci} A. Tredicucci, Y. Chen, V. Pellegrini, M B\"{o}rger, L. Sorba, F.
Beltram, and F. Bassani, ``Controlled Exciton-Photon Interaction in Semiconductor Bulk Microcavities'', Phys. Rev. Lett. {\bf 75}, 3906 (1995).

\bibitem{AFainsteinBulkGaAs} A. Fainstein, B. Jusserand, P. Senellart, J. Bloch, V. Thierry-Mieg, and R. Planel, ``Center-of-mass quantized exciton polariton states in bulk-GaAs microcavities'', Phys. Rev. B {\bf 62}, 8199 (2000).


% Optomechanics with DBR-based cavities

\bibitem{Trigo} M. Trigo, A. Bruchhausen, A. Fainstein, B. Jusserand, and V. Thierry-Mieg, ``Confinement of Acoustical Vibrations in a Semiconductor Planar Phonon Cavity'', Phys. Rev. Lett. {\bf 89}, 227402 (2002).

\bibitem{Anguiano} S. Anguiano, G. Rozas, A. E. Bruchhausen, A. Fainstein, B. Jusserand, P. Senellart, and A. Lema\^itre, "Spectra of mechanical cavity modes in distributed Bragg reflector based vertical GaAs resonators'', Phys. Rev. B {\bf 90}, 045314 (2014).

\bibitem{PSesin}P. Sesin, P. Soubelet, V. Villafa\~ne, A. E. Bruchhausen, B. Jusserand, A. Lema\^itre, N. D. Lanzillotti-Kimura, and A. Fainstein, ``Dynamical optical tuning of the coherent phonon detection sensitivity in DBR-based GaAs optomechanical resonators'', Phys. Rev. B {\bf 92}, 075307 (2015).

% Ps acoustics

\bibitem{Thomsen} C. Thomsen, H. T. Grahn, H. J. Maris, and J. Tauc, ``Surface generation and detection of phonons by picosecond light pulses'', Phys. Rev. B {\bf 34}, 4129 (1986).

\bibitem{Bartels} A. Bartels, T. Dekorsy, H. Kurz, and K. Koehler, ``Coherent Zone-Folded Longitudinal Acoustic Phonons in Semiconductor Superlattices: Excitation and Detection,'', Phys. Rev. Lett.{\bf 82}, 1044 (1999).


% P&P in microcavities

\bibitem{Kimura_coherentcavity} N. D. Lanzillotti-Kimura,  A. Fainstein, A. Huynh, B. Perrin, B. Jusserand, A. Miard, and A. Lemaitre, ``Coherent Generation of Acoustic Phonons in an Optical Microcavity'', Phys. Rev. Lett. {\bf 99}, 217405 (2007).

\bibitem{KimuraTheory} N. D. Lanzillotti-Kimura, A. Fainstein, B. Perrin, and B. Jusserand, ``Theory of coherent generation and detection of THz acoustic phonons using optical microcavities'', Phys. Rev. B {\bf 84}, 064307 (2011).

\bibitem{Kimura_doubleresonance} N. D. Lanzillotti-Kimura, A. Fainstein, B. Perrin, B. Jusserand, L. Largeau, O. Mauguin, and A. Lema\^itre, ``Enhanced optical generation and detection of acoustic nanowaves in microcavities'', Phys. Rev. B {\bf 83}, 201103(R) (2011).


% P&P in layered systems

\bibitem{Flor2012} M. F. Pascual-Winter, A. Fainstein, B. Jusserand, B. Perrin, and A. Lema\^itre, ``Spectral responses of phonon optical generation and detection in superlattices'', Phys. Rev. B {\bf 85}, 235443 (2012), and references therein.


% Supplementary Information


\bibitem{Johnson2002}
S.~G. Johnson, M.~Ibanescu, M.~A. Skorobogatiy, O.~Weisberg, J.~D. Joannopoulos, and Y.~Fink, ``Perturbation theory for maxwell's equations with shifting material boundaries'', Phys. Rev. E \textbf{65}, 066611 (2002).

\bibitem{Ding2010}
L.~Ding, C.~Baker, P.~Senellart, A.~Lema\^itre, S.~Ducci, G.~Leo, and I.~Favero, ``High frequency {G}a{A}s nano-optomechanical disk resonator'', Phys. Rev. Lett. \textbf{105}, 263903 (2010).

\bibitem{Florez2016}
O.~Florez, P.~F. Jarschel, Y.~A.~V. Espinel, C.~M.~B. Cordeiro, T.~P.~M. Alegre, G.~S. Wiederhecker, and P.~Dainese, ``Brillouin scattering self-cancellation'', Nature Communications \textbf{7}, 11759 (2016).

\bibitem{Eryiit1996}
R.~Eryi{\u{g}}it and I.~P. Herman, ``Lattice properties of strained {GaAs}, {S}i, and {G}e using a modified bond-charge model'', Physical Review B \textbf{53}, 7775--7784 (1996).

%\bibitem{SuppInfo} See the supplementary material for measurements and discussion related to differential reflectivity measurements performed in the GaAs epilayer.



\end{thebibliography}
\end{document}